# Indications of an Unmodelled Component in Spectrographic Measurements of Local Stars

## Charles Francis[1], Erik Anderson[2]

[1] 25 Elphinstone Rd., Hastings, TN34 2EG, UK.
[2] 679 Roca St., Ashland, OR 97520, USA.

ABSTRACT

*Context:* While CDM models and MOND give explanations for flat rotation curves of other galaxies, both present observational problems and the local gradient of the Milky Way's rotation curve is not flat.

*Aims:* We consider whether flat rotation curves could be an artifact of an unmodelled component in spectral shift.

*Methods:* In the absence of astrometric determinations of radial velocity, we apply a statistical test on a population of 20 440 Hipparcos stars inside 300 pc with known radial velocities and with accurate parallaxes in the New Hipparcos Reduction.

*Results:* The test rejects the null hypothesis, *there is no systematic error in spectrographic determinations of heliocentric radial velocity*, with 99.95% confidence. In a separate test on metal-poor stars, we find tension between calculations of the orbital velocity of the Sun from three populations of halo stars inside and outside of a cone of 60° semi-angle from the direction of rotation. Tension cannot be removed with only systematic distance adjustments.

*Conclusions:* We conclude that the most probable explanation is an unmodelled element in spectrographic determinations of heliocentric radial velocity with a probable cosmological origin, and propose that this unmodelled component, rather than CDM or MOND, is responsible for the apparent flatness of galaxy rotation curves.

## 1 Introduction

### 1.1 Background

According to conventional analyses, based on Doppler shifts of interstellar HI and CO, the slope of the Milky Way's rotation curve is close to zero over a wide range of distances (e.g. Combes 1991). This is normally accounted for by hypothesizing a cold dark matter halo (CDM), or by modifying gravity (MOND; Milgrom, 1994). Francis and Anderson (2009a, hereafter FA09a) calculated the local slope of the circular speed curve from the motions of local stars, finding a slope of -9.3 ± 0.9 km s$^{-1}$ kpc$^{-1}$, in agreement with the curves given by Combes (1991) found from CO and HI emissions (figure 1). This slope is at odds with the flat rotation curve due to a CDM halo or MOND, but agrees with the expected slope based on observed stellar and gaseous mass under Newtonian gravity.

Neither CDM nor MOND is free of empirical problems, and no explanation is found for either in fundamental physics. A full review goes beyond the scope of this paper, but a brief discussion of issues with CDM and MOND are given in section 1.2 and section 1.3. Our knowledge of stellar kinematics and cosmology is almost entirely dependent on spectroscopy, but there is no direct test of the Doppler law for objects at stellar distances. It is appropriate to consider a less radical departure from conventional physics, that an unmodeled component of spectral shift could offer an explanation for the slope of the Galactic rotation curve.

Such an unmodelled component would constitute new physics, but it might be suspected that new physics should come from the as yet unknown unification of quantum theory with general relativity and it is of fundamental importance to the scientific method that all possible explanations for empirical phenomena are considered. An unmodelled blueshift has been observed in signals from Pioneer, normally treated as an unmodelled acceleration (Anderson et. al., 2002). An unmodelled component of spectral shift which replicates MOND would modify heliocentric velocity, but not the component of radial velocity due to the Earth's orbital motion around the Sun (MOND is not seen in the solar system; an annual fluctuation in radial velocity measurements would be easily detectable).

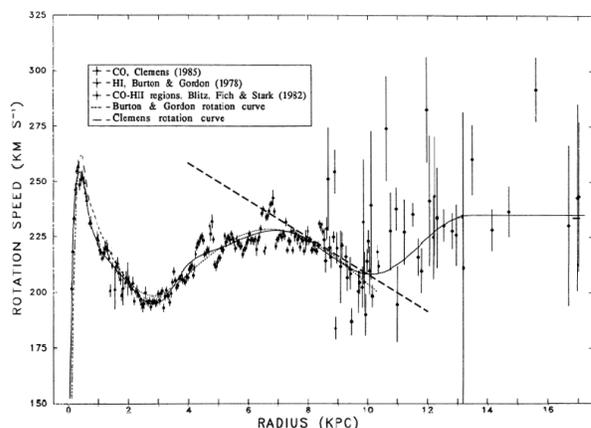

*Figure 3* Comparison between several rotation curves obtained by the method of terminal velocity in CO and HI emissions inside the solar radius, and by CO complexes associated with HII regions of known distances (from Clemens 1985). The IAU (1985) galactic constants have been adopted: $R_0 = 8.5$ kpc, $V_0 = 220$ km s$^{-1}$.

**Figure 1:** The Milky Way rotation curve from CO and HI (Combes, 1991), with superposed the gradient (dashed) from local stars (FA09a), adjusted to $R_0 = 8.5$ kpc, $V_0 = 220$ km s$^{-1}$ as used by Combes.

Astrometric determination of heliocentric radial velocity for individual stars will require data at least an order of magnitude more precise than current parallax measurements (Dravins et al, 1999). Data of this accuracy will be provided by Gaia. Here we consider two straightforward statistically based analyses using recently published stellar catalogues providing kinematically complete data for local stars. The results support the conclusion that the Galaxy rotation curve is not flat as described by CDM and MOND, but merely appears flat because an unmodelled component of spectral shift overstates heliocentric radial velocity.

Although a statistical test can never be 100% conclusive, and cannot reveal the cause of a correlation, we have considered numerous potential explanations (section 3.4) suggested by ourselves and by others, and we have not found an explanation in classical physics which could explain the outcome. In view of the fact that it is currently impossible to establish by empirical means the Doppler relation over astronomical distances, and in view that there is no accepted theoretical model of quantum cosmology, we should remain open to the prospect that phenomena described by CDM or MOND may be accounted for by the re-interpretation of redshift.

In section 2 we describe our stellar population for the first test, described in section 3. Section 3.2 and section 3.3 give the results. Section 3.4 considers possible causes of bias which could affect test results. Section 4 describes the second test, using a separate population of halo stars. The results are summarized in section 5. Section 6 contains a brief discussion of implications and suggests a possible cosmological explanation for the test results. Conclusions are summarized in section 7.

### 1.2    Issues for CDM

CDM is widely acknowledged as explaining galaxy rotation curves and other observations in astronomy and cosmology. However, CDM does not give an explanation as to why precisely the same acceleration law should be found in galaxies of many sizes and types. There is still no satisfactory theory of CDM in particle physics and CDM has not been discovered in earth based experiments. The CDM model is distressed by studies of globular clusters by Scarpa et al (2003, 2007, 2010), who find that a MONDian curve is obeyed although the amount of CDM in clusters is negligible, and by the dynamical studies of three elliptical galaxies based on the measurement of radial velocities of a large number of planetary nebulae by Romanowsky et al. (2003). They find evidence for "little if any dark matter in these galaxies" and conclude that this does not naturally conform with the CDM paradigm.

There is a substantial literature on lensing profiles indicating that the standard CDM model is not consistent. According to evolutionary models dark matter halos should have steep central density cusps (e.g., Navarro, Frenk, & White 1997) but they appear not to (e.g., de Blok, Bosma & McGaugh 2003; Swaters et al. 2003). In a survey of about 3,000 galaxies, Biviano & Salucci (2006) find that X-ray determination of the baryonic component of dark matter halos fits evolutionary models, but subhalo components do not. Martel and Shapiro (2003) have examined the profile of lenses for a number of evolutionary models. While they find quan-titative fits for many properties, they find that the models do not correctly reproduce the central region. Park and Ferguson (2003) studied the lensing produced by Burkert halos and found, "For the scaling relation that provides the best fits to spiral-galaxy rotation curve data, Burkert halos will not produce strong lensing, even if this scaling relation extends up to masses of galaxy clusters. Tests of a simple model of an exponential stellar disk superimposed on a Burkert-profile halo demonstrate that strong lensing is unlikely without an additional concentration of mass in the galaxy center (e.g. a bulge)".

Power et al (2003) comment on discrepancies between analytic models and halo distribution required by galaxy rotation curves. In particular they state "there is no well defined value for the central density of the dark matter, which can, in principle, climb to arbitrarily large values near the centre". Of this result they say "there have been a number of reports in the literature arguing that the shape of the rotation curves of many disk galaxies rules out steeply divergent dark matter density profiles" and conclude that it "may signal a genuine crisis for the CDM paradigm on small scales".

The dynamical mass in galaxy clusters is typically about a factor of 4 or 5 larger than the observed mass in hot gas and in the stellar content of the galaxies. This rather modest discrepancy viewed in terms of dark matter has been called the baryon catastrophe (White et al 1993); there is not enough non-baryonic dark matter in the context of standard CDM cosmology.

### 1.3    Problems for MOND

MOND has been successful in a range of contexts, and exists in a number of flavours, such as TeVeS (Tensor-Vector-Scalar gravity, Bekenstein, 2004) and NGT (nonsymmetric gravitational theory, Moffat, e.g., 2005) which propose relativistic models replacing exotic matter with exotic gravity, but the physical basis of these models remains obscure and the division between the MONDian and Newtonian regimes appears artificial.

MOND has a problem in galaxy clusters where central accelerations higher than those predicted have been found from studying temperature gradients (The & White, 1988) and in modeling Ly-α absorbers (Aguirre et al., 2001).

The lower galactic masses of no-CDM theories do not account for the amount of lensing in all cases. Zhao et al. (2006) tested lensing in Bekenstein's relativistic MOND (TeVeS), and concluded that lensing may be a good test for CDM theories. They found that "TeVeS succeeds in providing an alternative to general relativity in some lensing contexts; however, it faces significant challenges when confronted with particular galaxy lens systems". In studies on the Bullet cluster, 1E0657-558 Clowe et. al (2004, 2006) have shown that MOND does not account for lensing without non-baryonic dark matter.

## 2    Our Stellar Population

### 2.1    Stellar Databases

To minimize the influence of random errors on results, it is important to use stars for which accurate measurement is available. Hipparcos provided parallax measurements of unsurpassed accuracy. We derived a stellar population with kinematically complete data by combining astrometric parameters from the recently released catalogue, *Hipparcos, the New Reduction of the Raw Data* (van Leeuwen, 2007a; hereafter "HNR") plus the Tycho-2 catalogue (ESA, 1997) with heliocentric radial velocities contained in the *Second Catalogue of Radial Velocities with Astrometric Data* (Kharchenko, et al., 2007; hereafter "CRVAD-2").

Systematic parallax errors in the original Hipparcos catalogue are stated at less than 0.1 mas (ESA, 1997), or less than 3% for a star at 300 pc. HNR claims improved accuracy by a factor of up to 4 over the original Hipparcos catalogue (ESA, 1997) for nearly all stars brighter than magnitude 8. The improvement is due to the increase of available computer power since the original calculations from the raw data, to an improved understanding of the Hipparcos methodology, which compared positions of individual stars to the global distribution and incorrectly weighted stars in high-density star fields leading to the well-known 10% error in distance to the Pleiades, and to better understanding of noise, such as dust hits and scan-phase jumps. *Validation of the New Hipparcos Reduction* (van Leeuwen, 2007b) "confirms an improvement by a factor 2.2 in the total weight compared to the catalogue published in 1997, and provides much improved data for a wide range of studies on stellar luminosities and local galactic kinematics."

CRVAD-2 contains most of the stars in two important radial velocity surveys: *The Geneva-Copenhagen survey of the Solar neighbourhood* (Nordström, et al., 2004; hereafter "G-CS"), which surveyed nearby F and G dwarfs, and *Local Kinematics of K and M Giants from CORAVEL* (Famaey et al., 2005; hereafter "Famaey"). We included about 300 stars in G-CS and Famaey not given in CRVAD-2 and incorporated the revised ages for G-CS II (Holmberg, Nordström and Andersen, 2007).

We restricted the populations to stars for which standard parallax errors were less than 20% of the quoted parallax. A distance cut of 300 pc was also applied. After the distance cuts, the populations contained very few stars with large motion errors. The accuracy of proper motions in HNR is better by a factor of about two than that of Tycho-2, which compared star positions from the Hipparcos satellite with early epoch ground-based astrometry. We used a mean value from HNR and Tycho for proper motion, inversely weighted by the squared quoted error, to obtain the best possible figure. The mean error in transverse velocity is 0.34 km s$^{-1}$, about 1% of the mean transverse velocity, 32.9 km s$^{-1}$. The mean error in heliocentric radial velocity for the population is 1.3 km s$^{-1}$, for stars also in G-CS the error is 0.87 km s$^{-1}$, and for stars also in Famaey it is 0.26 km s$^{-1}$. These random errors are of an order of magnitude less than the systematic error required to explain the results of the regression test.

## 2.2 Selection Criteria

Our population of 20 440 stars is obtained by applying the following selection criteria:

(i) Heliocentric distance within 300 pc based on HNR parallaxes and parallax error less than 20% of parallax (see section 2.3).

(ii) Heliocentric radial velocity given in CRVAD-2, GC-S or Famaey and uniquely identified to a Hipparcos catalogue number. CRVAD-2 figures were used by default, as CRVAD-2 gives a weighted mean for stars in Famaey having radial velocities from other sources. We excluded stars for which no radial velocity error was given, or for which the quoted error was greater than 5 km s$^{-1}$.

(iii) The object is either a single star or a spectroscopic binary with a computed mean radial velocity. This criterion is determined from flags provided by G-CS, Famaey, Tycho-2, and CRVAD-2.

(iv) It is usual in statistical analyses of data to eliminate outliers more than three (or fewer) standard deviations from the mean, because outliers tend to have a disproportionately large affect on results. This cannot be done here because the distributions are far from Gaussian and contain a high proportion of fast moving stars. Velocities opposing any error in the mean will be preferentially removed, resulting in a compounded error and leading to non-convergence on iteration of the method. It remains important to remove stars with extreme velocities, especially those with contrary orbits or with orbits excessively inclined to the Galactic plane. A more disperse distribution was found for stars aged over 10 Gyrs. We applied a cut on stars with velocities outside of an ellipsoid,

$$\frac{(U+12)^2}{200^2} + \frac{(V+42)^2}{200^2} + \frac{(W+7)^2}{120^2} < 1 \,, \qquad (2.2.1)$$

corresponding approximately to a 4 s.d. cut on each axis for the population of old stars, and to over 6 s.d. for the remaining population. This removed 86 stars.

A number of kinematic studies have concentrated on stars in open clusters. These are likely to be over-represented in CRVAD-2. Even if they were not over-represented the existence of groups localised in 6 dimensional phase space would affect the regression test, because it would mean that bins are not kinematically independent. We removed 138 candidate stars from the Hyades cluster (by position in phase space), 46 from Alpha Persei (Makarov 2006), 27 from the Pleiades (Makarov 2002), 8 from Praesepe (Patience et al., 2002), and 3 from the Coma star cluster (Mermilliod et al., 2008).

## 2.3 Parallax Errors

Because parallax distance is measured as an inverse law of parallax angle, errors are not symmetrical and a systematic distance error is introduced (this is a part, but not the main part, of the Lutz-Kelker bias which concerns estimates of absolute magnitude; Lutz & Kelker, 1973, 1974, 1975). For example, for two measurements with 20% error above and below the true parallax, $\pi$, of a given star, the mean parallax distance, $R$ pc, is given by

$$R = \left(\frac{1000}{\pi(1-0.2)} + \frac{1000}{\pi(1+0.2)}\right) \div 2 = \frac{1000}{\pi(1-0.2^2)} \,, \qquad (2.3.1)$$

giving a mean error of +4%. For a Gaussian error distribution with $\sigma = 20\%$ of $\pi$, we calculate an expected systematic error of +1.6% (by numerical solution of the integral). Over 70% of the stars in the population have parallax errors less than 10%. The systematic error goes as the square of the random error and can be estimated at below 1%. We compensated using a pragmatic approximation,

$$R = \frac{1000}{Plx}(1 - 0.4(ePlx/Plx)^2) \qquad (2.3.2)$$

where $Plx$ and $ePlx$ are the measured parallax and parallax error given in HNR.

## 2.4 Kinematic Bias

G-CS and Famaey are deemed to be free from kinematic selection bias. The remaining radial velocities in CRVAD-2 are derived from the *General Catalog of Mean Radial Velocities* (Barbier-Brossat and Figon, 2000) and the *Pulkovo Catalog of Radial Velocities* (Gontcharov, et al., 2006). These are compilations from various sources. Binney et al. (1997) have claimed that such compilations should not be used in kinematic studies because they contain a bias toward high proper motion stars. However, they did not give a statistical analysis for their conclusion, but justified it from a graph (their fig. 2) with a logarithmic scale which exaggerates evidence of bias by two orders of magnitude. In fact a bias towards high proper motion will result if, as may be expected, near stars are chosen in surveys. Bias toward high proper motion is, in itself, not evidence of a velocity bias. Binney et al also used a sample of stars near the South pole. Because of the angle of the Earth's axis, and the motion of the Sun, this will also introduce bias in proper motions.

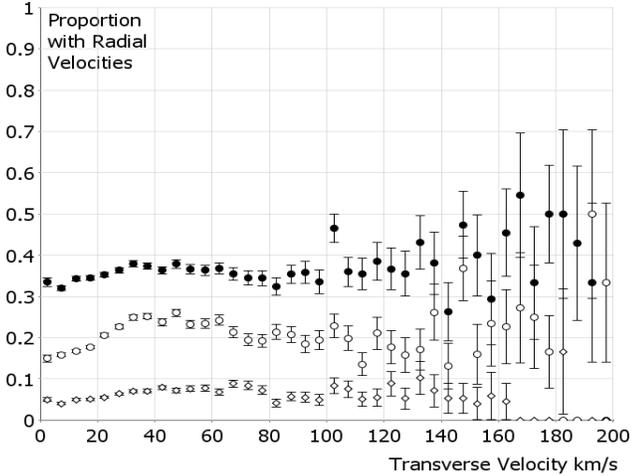

**Figure 2:** The ratio of the number of stars in each bin with known radial velocity to the number of stars in each bin for the whole population. The whole sample is shown with filled circles. Stars in G-CS are shown as open circles. Stars in Famaey are shown as open circles.

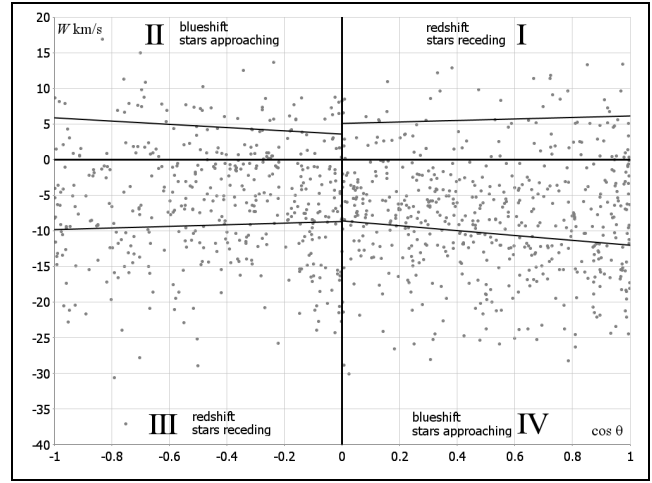

**Figure 3:** Regression of $W$ for stars with $0.21 < B - V \leq 0.32$ mag plotted against the cos of the angle subtended to the $W$ axis, showing four "passes". Correlations are low but the total number of quadrants with absolute value of the component of velocity increasing with the absolute value of the cosine is significant.

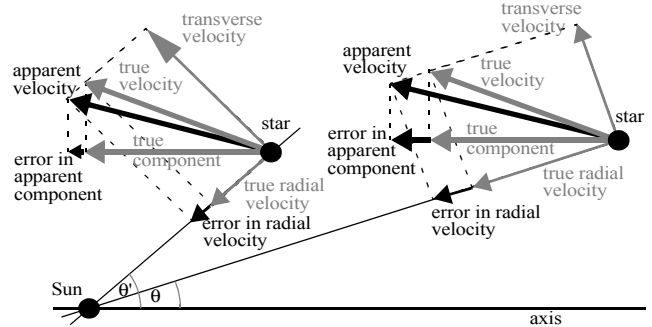

**Figure 4:** For stars with equal true velocity and different positions, a systematic proportionate error in radial velocity would result in a greater error in the component of velocity along a given axis (horizontal) for the star subtending a narrower angle with that axis.

After restricting the sample to stars with radial velocities available to Binney et al. it is still not possible to reproduce the level of bias which they found. Binney et al. used a sample limited by luminosity distance. In view that Hipparcos provides accurate parallax distances for nearby stars, this is a very strange decision. Because they did not give details of their distance model, it is not possible to reproduce their result, but it is possible to determine that Binney et al.'s luminosity distance model is responsible for the major part of the bias in their sample. The reported selection bias was introduced by the authors themselves.

To test for a selection bias, we bin the stars by transverse velocity, and plot the ratio of the number for which we have radial velocities to the number of stars in each bin for the whole population (figure 2). A perfectly flat line is not expected because a magnitude limited sample will contain dependencies on stellar type (e.g FA09a; Dehnen, 1998). There is no evidence of a kinematic bias which would affect the results of the test described here.

## 3 Regression Test

### 3.1 Method

A statistical test of stellar motions to detect an unmodeled component in spectrograph radial velocity must overcome considerations due both to noise and to structure. The velocity dispersion in the local population is much greater than the suggested error in heliocentric radial velocity which would account for the flattening of rotation curves. Even without moving groups, the distribution of phase space in the Solar neighbourhood is highly structured in the Galactic plane (FA09a and references cited therein), so that it is not possible to directly compare mean motions in one direction with another. A test is required which will not be affected by the structure of the velocity distribution, real velocity gradients or bulk streaming motions.

As a spiral galaxy, the Milky Way has a highly structured stellar velocity distribution (Francis & Anderson, 2009b, hereafter FA09b). Stars move along an arm from before apocentre to after pericentre, and cross the other arm close to the semi-latus rectum. It is to be expected that there will be a small velocity gradient, and a small velocity divergence among outgoing stars (Hyades stream), and a small net velocity convergence among ingoing stars in the arm (Alpha Ceti stream). The current position of the Sun, near the inside of the arm, means that we are close to stars near pericentre (Alpha Lacertae and Sirius streams). A velocity gradient may be expected across the arm. Further irregularities can be expected from regions of star formation, which create stars in more nearly circular motions and typically close to apocentre (Pleiades stream).

To test for a signature in such a noisy distribution we binned the population of 20 440 stars into 20 colour bins, each containing over 1 000 stars, and tested the velocity components, $U$, towards the Galactic centre, $V$, in the direction of Galactic rotation, and $W$, perpendicular to the Galactic plane. We did not divide bins into dwarfs and giants; there is no particular dynamical reason for the binning strategy (random bins were also used, with no significant difference in results, but we report on the use of colour bins for reason of reproducibility by other researchers).

The component of velocity, $v_{axis}$, in the direction of the axis was plotted against the cosine of the angle, $\theta$, subtended by the star with that axis (figure 3). An error in heliocentric radial velocity will contribute more to $v_{axis}$ for stars which subtend a narrow angle with the axis (figure 4), and will tend to generate a correlation between $v_{axis}$ and cos θ in each of the four quadrants. Any correlation will be low because of the high degree of scatter. Using the cosine linearizes the distribution of the population on the abscissa. The four quadrants of the plot represent stars positioned in either direction along the axis (quadrants I & IV opposed to quadrants II & III), and stars approaching (quadrants II & IV) and receding (quadrants I & III).

Under the null hypothesis, that *there is no systematic error in spectrographic determinations of heliocentric radial velocity*, there should be a 50-50 split of plots with absolute component of velocity increasing or decreasing with abs(cos θ). Trials showing increasing abs($v_{axis}$) with abs(cos θ) were designated "passes" for the alternate hypothesis, *Heliocentric radial velocities are overstated*. The confidence limit for the alternate hypothesis is calculated in the standard manner from the cumulative binomial distribution with probability 0.5. In order to account for the flattening of the rotation curve an overstatement of velocity would be required in the direction *V* of orbital motion, but not in the radial, or *U*, direction.

For typical orbits in the disc, motion perpendicular to the disc may be treated as an independent oscillation superimposed on an orbit in the plane of the disc (since the oscillation is perpendicular to the centripetal force and to orbital motion). Stars with greater *W*-velocities will oscillate at greater amplitude, and tend to be further from the central plane of the disc, and a 50-50 split cannot be expected (this was confirmed with a simple numerical model). We therefore ignored results on the *W*-axis.

The use of a test for the component of motion along a single axis eliminates gradients and other irregularities perpendicular to that axis. A simple binary test eliminates any weighting due to the number of stars in each quadrant. Using quadrants with opposite directions of space and motion, and using a test which does not depend on stellar distances, effectively eliminates gradients in the direction of the axis, while the split into stars approaching and receding eliminates effects due to velocity divergences (any bias due to a velocity divergence in the quadrant for stars approaching from one direction would be cancelled by an opposite bias for stars receding in the opposite direction).

### 3.2    Main Result

The overall result from 80 quadrants on the *V*-axis was 49 passes. This is significant at 97.2%. Because outliers in regression have a disproportionate effect on results, it is normal to restrict the population to within 3 or fewer standard deviations of the mean. FA09a found the velocity ellipsoid,

$$\frac{(U+11.2)^2}{70^2} + \frac{(V+13.6)^2}{40^2} + \frac{(W+6.9)^2}{23^2} < 1 \;.  \qquad (3.2.1)$$

This ellipsoid contains 14 794 stars and represents the bulk of stars with thin disc motions. The velocity ellipsoid has no dependency on space coordinates, so does not introduce truncation bias under the null hypothesis. After restricting to this velocity ellipsoid, the number of passes out of 80 rose to 55, leading to rejection of the null hypothesis with 99.95% confidence. This ellipsoid has approximately 3σ semi-axes.

The result of 28 passes from 80 trials on the *U*-axis rejects the alternate hypothesis with 99.75% confidence, and shows that the overall result is not due to a systematic understatement of Hipparcos parallax distance which would affect the *U*- and *V*-directions equally.

### 3.3    Ancillary Results

After restricting the population to stars with less than 10% parallax errors there are 50 passes out 80 trials on the *V*-axes for the whole population and 52 passes for stars in the velocity ellipsoid, rejecting the null hypothesis with 99.5% confidence.

| Axis | region | trials | passes | confidence |
|------|--------|--------|--------|------------|
| *U* | All | 80 | 28 | - |
| *V* | All | 80 | 49 | 97.2% |
| | | | | |
| *U* | Ellipsoid | 80 | 33 | - |
| *V* | Ellipsoid | 80 | 55 | 99.95% |

**Table 1:** Components of radial velocity are overstated in the direction of orbital motion, but not in the direction of motion radial to the Galactic centre.

| Axis | region | I | II | III | IV |
|------|--------|---|----|-----|----|
| *U* | All | 7 | 8 | 10 | 3 |
| *V* | All | 14 | 11 | 11 | 13 |
| | | | | | |
| *U* | Ellipsoid | 7 | 7 | 10 | 9 |
| *V* | Ellipsoid | 13 | 14 | 14 | 14 |

**Table 2:** The number of passes for each axis out of twenty trials in each quadrant for the entire population and in the velocity ellipsoid.

To reduce the effect of possible extinction, we restricted the population to 100 pc. This leaves a small population in certain quadrants, which would be expected to reduce the significance of any correlation under the alternate hypothesis, and so reduce and the expected pass rate. It does not affect probabilities or the validity of the test under the null hypothesis. In one quadrant on the *V*-axis there is only one star and no test is possible. The result from 79 trials on the *V*-axes was 79 passes for the whole population, and 53 passes within the velocity ellipsoid, rejecting the null hypothesis with 99.7% confidence.

To test whether the results might be caused by a velocity gradient in *U*- or *V*- velocity components in a direction perpendicular to the galactic plane we restricted the population to stars within 50 pc from the Sun in the *W*-direction, substantially less than the scale height of the thin disc, which is between 250 and 300 pc (Lopez-Corredoira et al, 2002; Kent et al., 1991). This made little difference to results, giving 25 passes on the *U*-axis for the whole population, and 31 passes within the ellipsoid. For the *V*- axis there were 47 passes fro the whole population and 52 passes for stars within the velocity ellipsoid, which is significant at 99.5% confidence.

Clusters are localised in space as well as motion, and were removed from the population because many stars would appear in the same part of a given quadrant for a number of different bins.Although this would not bias the test, it would reduce its significance my increasing the likelihood that different bins produce the same result. Moving groups and OB associations are more spread in space and had not been removed, but they may have some directional dependency which could potentially affect the result. We removed stars from the Upper Scorpius Association, Upper Centaurus Lupus Association, Lower Centaurus Crux Association, Vela OB2 Association, Trumpler 10 Association, Collinder 121 Association, Perseus OB2 Association, Alpha Persei Association, Cassiopeia-Taurus Association, Lacerta OB1 Association, Cepheus OB2 Association, Cepheus OB6 Association, (De Zeeuw et al., 1999), Tucana / Horologium Moving Group (Zuckerman, et al., 2001a; Song et al., 2003), AB Doradus Moving Group (Zuckerman, et al., 2004), Ursa Major Moving Group (Soderblom et al., 1993), Beta Pictorus Moving Group (Zuckerman, et al., 2001b; Song et al., 2003), Castor Moving Group (Barrado y Navascues et al., 1998), Carina Near Moving Group (Zuckerman, et al., 2006), TW Hydrae Moving Group (Reid, 2003). The result from 80 trials on the $V$-axes remained at 49 passes for the whole population and rose to 56 passes, for the velocity ellipsoid, rejecting the null hypothesis with 99.97% confidence.

### 3.4 Possible Sources of Bias

It is important in a test such as this to consider every possible cause of bias.

*Measurement Errors:* Quoted measurement errors in proper motion, radial velocity, and parallax distance are of the order of 1%. The pass rate in the velocity ellipsoid on the $V$-axis was still significant at 95% after a 5% systematic increase in radial distance, and hence in transverse velocity. Systematic measurement errors are also excluded because the result on the $U$-axis does not match that on the $V$-axes.

*Truncation Bias:* The velocity ellipsoid is independent of position in space, and hence is independent of angle to the chosen axis. The velocity ellipsoid therefore does not introduce a truncation bias. The reversed correlation on the $U$-axis excludes the possibility that the result is due to truncation bias arising from the fact that measurement errors in heliocentric radial velocity are slightly larger (<1 km s$^{-1}$) than those in transverse velocity; the division of the population into stars approaching and stars receding cuts stars which cross the horizontal axis because of measurement errors, and could potentially produce a bias towards passes. In fact, only a few stars could be affected, and estimates of the magnitude of this bias show that it is greatly outweighed by random factors in the motions of stars.

*Colour Bias:* A systematic error at given colour arising from gravitational redshift and/or convective blueshift at the stellar photosphere would affect all axes in the same way, and would produce opposite results in quadrants for stars approaching and receding. This would not increase the net pass rate under the null hypothesis, and can thus be eliminated as a possible cause of the test result.

*Deviation Due to Orbital Motion:* The test result shows a high proportion of passes in the direction, $V$, of orbital motion. It may be expected that there will be a small drop in heliocentric radial velocities due to orbital curvature. This might be of the order of 1% for a star at a distance of 100pc, an order of magnitude less than the error required to account for the effect. In any case a drop in heliocentric radial velocity would favour the null hypothesis.

*Stellar Streams:* The velocity distribution of local stars is highly structured and is dominated by bulk streaming motions (e.g. Eggen, 1958; Dehnen 1998; Famaey et al., 2005). FA09b showed that streams are caused by the gravitational alignment of orbits leading to the spiral structure of the Milky way. To first order, stellar streams do not affect the outcome of the test, because stream velocity is independent of position. If, in a fast moving stream, stars whose position is perpendicular to stream motion are preferentially selected over those whose position is in line with stream motion (e.g. due to bias toward high proper motions) then the regression test could produce a low proportion of passes on an axis parallel to stream motion, and a high proportion on an axis perpendicular to stream motion. The Sirius stream has mean direction roughly parallel to the $U$-axis, but is slow moving with respect to the Sun (e.g.

FA09a) and would produce an opposite result. The Hyades, Pleiades and Hercules streams have mean direction about midway between the $U$-and $V$-axes. There thus does not appear to be a mechanism according to which streaming motions could lead to the result.

*Velocity gradients:* The model of spiral structure predicts a velocity gradient across the width of a spiral arm (stars near pericentre will be found near the inner rim of the arm). However, because the test is independent of direction perpendicular to the axis, its outcome is not affected by a velocity gradients perpendicular to the axis. Because separate tests are used in each quadrant of the plot, a real velocity gradient in the direction of the axis would increase the pass rate in one quadrant and would decrease the pass rate in the opposite quadrant. This would not be expected to alter the 50% pass rate under the null hypothesis. In practice the results for individual quadrants (table 2) do not show preference for one quadrant over another.

*Velocity divergences:* A small divergence may be expected in the population of stars crossing the arm on the outward part of their orbit (Hyades stream) and a small convergence may be expected among stars following the arm on the inward part of their orbits (Alpha Ceti & Sirius streams), but any bias due to a velocity divergence in the quadrant for stars approaching from one direction would be cancelled by an opposite bias for stars receding in the opposite direction. Velocity divergences could not be responsible for the test result.

*Extinction and other irregularities:* Stars in open clusters were removed from our population, but other localised irregularities, such as moving groups or regions of extinction due to dust clouds will not alter the prediction of a 50% pass rate under the null hypothesis, but they would affect the significance of the result if they caused a quadrant to repeat results in different bins. The removal of moving groups and OB associations made no important difference to test results. To eliminate extinction we reduced the cutoff distance to 100pc. This lead to smaller samples but the result remained significant at 99%. If the result were caused by local irregularities, it would show in particular quadrants, but a breakdown of results by quadrant showed no such preferences.

## 4 Solar Motion Relative to the Halo

### 4.1 Population

Kinematics of metal-poor stars in the Galaxy (Beers et al., 2000) is a catalogue of Galactic stars with metal abundances in the range $-4 \le [Fe/H] \le 0.0$ . These are halo and thick-disc stars with much higher velocities than thin-disc stars. Radial velocities, photometric distances, and proper motion data are provided from a variety of high-precision sources for 1 258 stars. We selected a population of 545 mainly halo stars with $[Fe/H] < -1.5$ , containing subpopulations of 207 dwarfs, 216 giants, and 122 RR Lyrae stars. The results of the regression test on this population give extremely high proportions of passes, but may not be valid because the Lutz-Kelker bias means that photometric distances are subject to high systematic errors. We tested this population using a calculation of solar motion relative to the halo which allows corrections for systematic error in both distance and radial velocity.

### 4.2 Method

For [Fe/H]<-1.5, the metal-poor halo is essentially a non-rotating spheroid (Gilmore, Wyse & Kuijken, 1989). To confirm this result, and to eliminate any residual rotation from thick disc stars, we applied a variable cut on $W$-velocity, for $n = 0, ..., 10$, $[W + 7] > 10n$, and plotted mean $V$-velocity against the $W$-cut (figure 4). If there is a net rotation in the halo, we should expect this to be more apparent for stars whose orbits have lower inclination to the disc. In practice, the plots for $V$-velocity against cut on $W$ level off above about $50\,\mathrm{kms}^{-1}$, at which point thick disc stars are eliminated. This confirms Gilmore et al.'s result, that rotation is effectively eliminated for $W$-velocities over $50\,\mathrm{kms}^{-1}$ and metalicities $[\mathrm{Fe/H}] < -1.5$. Stars of different types in the halo would not be expected to have a different net rotation rate. In practice figure 4 shows poor agreement between the rate of rotation of RR Lyrae and the other populations.

We subdivided the populations using a cone with semi-angle $60°$ from the $V$-axis and calculated the mean velocity in the direction of Galactic rotation for the populations inside and outside the cone. Some selection effects are apparent. If the populations were evenly distributed across the celestial sphere then dividing them with a cone of semi-angle $60°$ would produce a 50-50 split of stars inside and outside the cone. We found a 53-69 split of the RR Lyrae population, a 55-161 split of giants and a 21-186 split of the dwarf population. However, there is no reason to think that a spacial bias will cause a kinematic bias or affect the analysis given here, beyond reducing the sizes of certain populations after the split, and thereby increasing uncertainty in results.

There should be no systematic difference between the mean $V$-velocity for populations inside and outside the cone, but for each subpopulation the calculated velocity for stars inside the $60°$ cone is greater than that of stars outside the cone (figure 5). Although the errors are of the order of $1\sigma$ and individually are not significant, the repetition of the pattern across three populations shows a systematic error which is significant.

### 4.3   Distance Adjustments

A systematic difference between velocities inside and outside the cone could be caused by a systematic understatement of distance or a systematic overstatement of heliocentric radial velocity. We applied systematic distance adjustments of 20% to dwarfs, 25% for giants, and 15% for RR Lyrae (figure 6), but this does not remove systematic differences between motion for stars inside and outside the cone. The resulting prediction of solar orbital velocity, $259 \pm 9\,\mathrm{kms}^{-1}$, is in poor agreement an estimate of solar orbital velocity of $225 \pm 5$ $\mathrm{kms}^{-1}$ found from measurement of the proper motion of SgrA* (Reid and Brunthaller, 2004), under the assumption that Sgr A* is stationary at the Galactic centre together with a combined estimate for $R_0$ of $7.45 \pm 0.17\,\mathrm{kpc}$ found from $7.2 \pm 0.7\,\mathrm{kpc}$ from $H_2O$ masers (Reid, 1993), $7.52 \pm 0.10\,(\mathrm{stat}) \pm 0.35\,(\mathrm{sys})$ from infrared photometry of bulge red clump stars (Nishiyama et al., 2006), $7.2 \pm 0.3\,\mathrm{kpc}$ from globular clusters (Bica et al., 2006), $7.62 \pm 0.32\,\mathrm{kpc}$ from Keplerian motions (Eisenhauer et al., 2005), and $7.6 \pm 0.4\,\mathrm{kpc}$ from statistical parallax of RR Lyrae stars (Layden et al., 1996).

### 4.4   Distance and Velocity Adjustments

We estimated from the mass models of Klypin, Zhao and Sommerville (2002) that, if the Galactic rotation curve is explained by a cosmological component of spectral shift (rather than by CDM or by MOND) then at the solar radius the cosmological component would contribute about 20-25% to spectral shift in the direction of orbital motion. This affects velocities inside the cone more than those outside of it. We applied a cut of 23% to radial velocities, although a strict proportional decrease is not indicated, because the

$U$-component should not be affected, it is reasonable to use an approximation. After applying this factor to radial velocity a good fit was obtained by increasing distances increasing distances to dwarfs by 7%, increasing distances to giants by 10%, and decreasing distances to RR Lyrae by 5%. (figure 7 & figure 8). The prediction for the solar orbital velocity with these adjustments, $221 \pm 7$ $\mathrm{kms}^{-1}$ is consistent with other estimates.

### 5   Results

We have applied tests which look for systematic differences between heliocentric radial velocity and transverse velocity. In section 3 we described a test which is largely independent of the actual distribution in phase space of local stars. It looks only for a correlation between the component of velocity on a given axis and the angle subtended by the star to that axis. We found no correlation on the $U$-axis, but highly significant correlations on the $W$- and $V$-axes, rejecting the null hypothesis that *there is no systematic error in spectrographic determinations of heliocentric radial velocity*, with 99.95% confidence.

In a separate test we calculated the solar motion in the direction of Galactic rotation relative to the halo. It is known that there is possible systematic bias (the Lutz-Kelker bias) in photometric distances, but we were unable to remove tension between results for different populations simply by making only systematic distance adjustments. After a systematic reduction in heliocentric radial velocities, based on the idea that galactic rotation curves may be distorted by a component of spectral shift due to cosmological expansion, tension between mean $V$ for different populations was easily removed by relatively small systematic distance adjustments.

### 6   Discussion

The cosmological redshift prediction of general relativity based on classical wave motions is clear, but general relativity does not consider the possibility that photons from astronomical objects should be described using quantum theory. Even in the absence of an accepted model of quantum gravity, we should consider the possibility that photons from distant astronomical bodies should be treated quantum mechanically, and we should recognize that if this is the case then the calculation of astronomical spectral shift goes beyond classical general relativity. In this case it is not possible, on theoretical grounds, to exclude the possibility that spectral shifts have a cosmological component in addition to the accepted Doppler component. If such a component were present it would present an alternative solution to the cosmological problems addressed by CDM and MOND. A rigorous test of this idea necessitates direct comparison of astrometric radial velocities with spectrographic radial velocities for individual stars. This will be possible for near, high velocity, stars with Gaia, but cannot be done at current astrometric precision.

Determination of spectral shift is straightforward, well established, and not in itself open to systematic measurement errors of the type seen in this paper. The results of the regression test cannot be accounted through systematic distance adjustments, because the correlation (if there is one) is in the direction radial to the Galaxy ($U$), and because the required systematic error is greater by an order of magnitude than the systematic error in Hipparcos. Velocity components are not expected to vary greatly with position over the distances of stars tested, and a simple velocity gradient could not in any case be responsible for the results because this would produce as many fails as passes. We have excluded truncation bias and bias due to moving groups and regions of extinction.

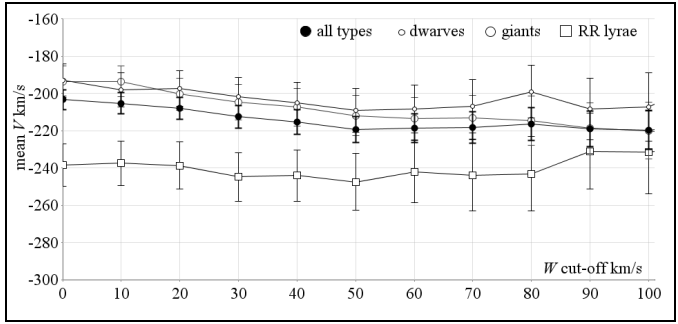

**Figure 4:** Average velocities in the direction of Galactic rotation of stars with metal abundances [Fe/H] $< -1.5$, unadjusted for systematic errors in distance or radial velocity. A cut is applied on $W$-velocity, shown as the horizontal axis, to eliminate rotational velocity seen by the rise in the plots for lower values of the cut. There is tension between the rate of rotation of RR Lyrae and the other populations.

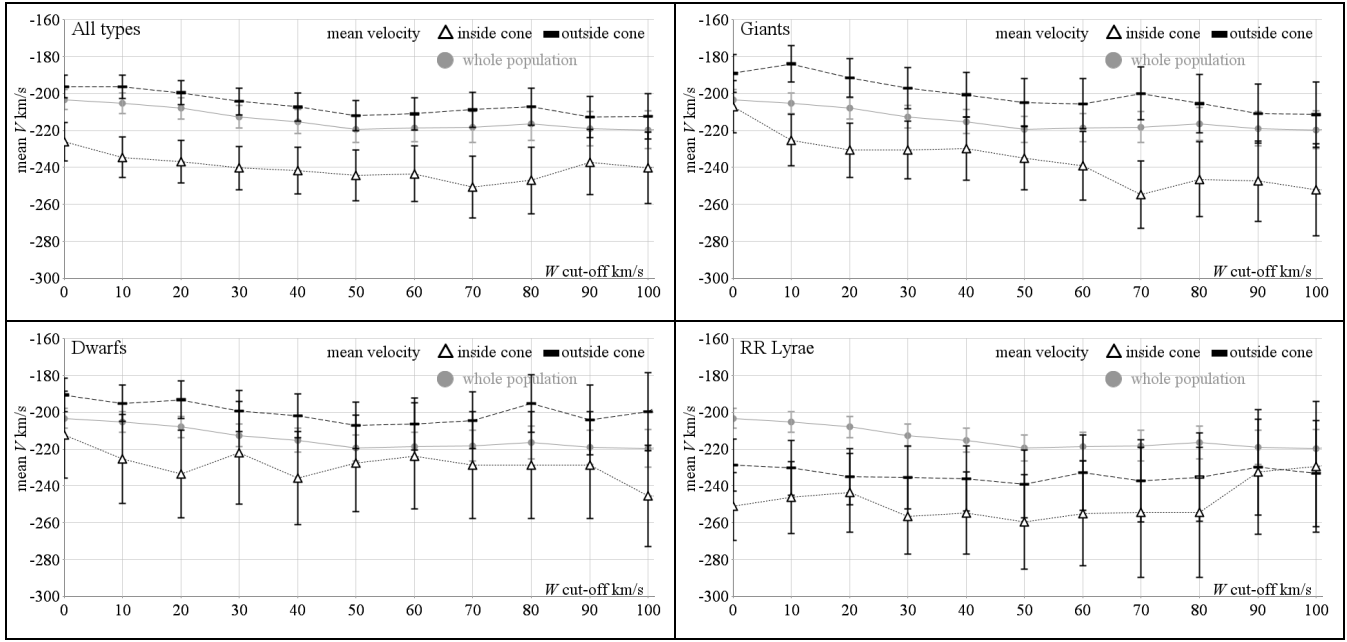

**Figure 5:** Mean velocities in the direction of Galactic rotation of stars with metal abundances [Fe/H] $< -1.5$, unadjusted for systematic errors in distance or radial velocity. A cut is applied on $W$-velocity, shown as the horizontal axis, to eliminate rotational velocity seen by the rise in the plots for lower values. Triangles show mean velocity within a 60° cone of the $V$-axis. Bars show mean velocity outside the cone. Grey circles show the mean for the whole population.

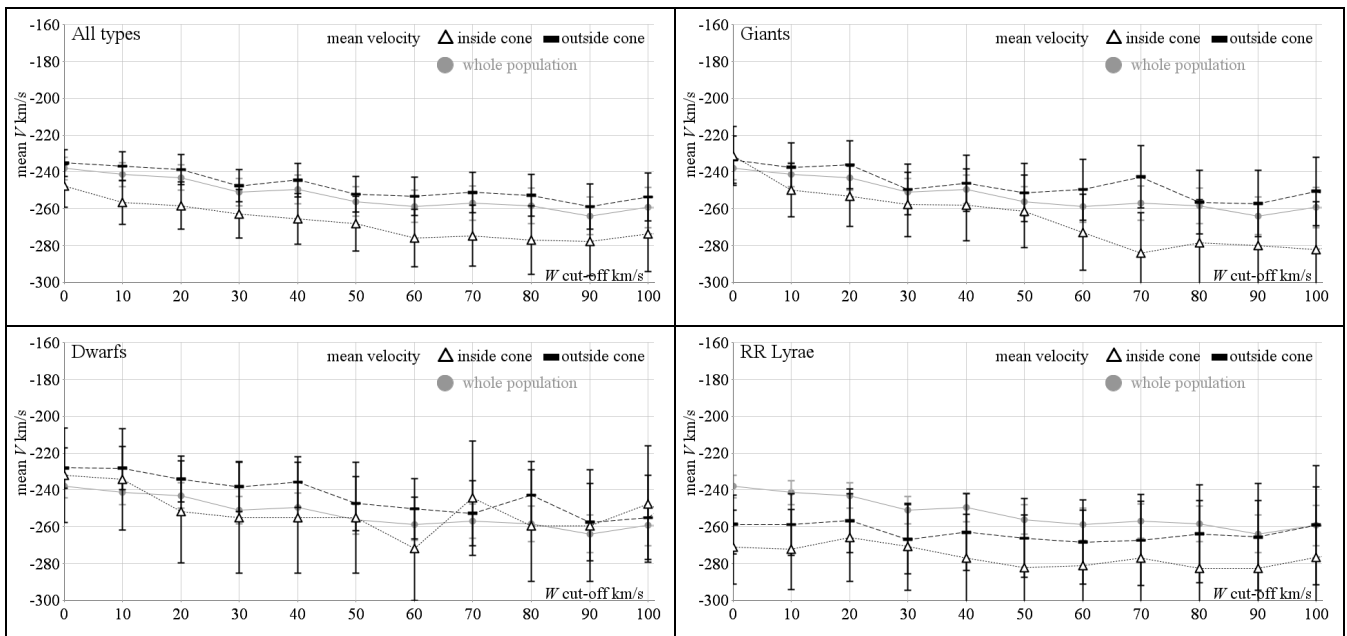

**Figure 6:** As figure 5, but with increases in distance of 20% for dwarfs, 25% for giants and 15% for RR Lyrae.

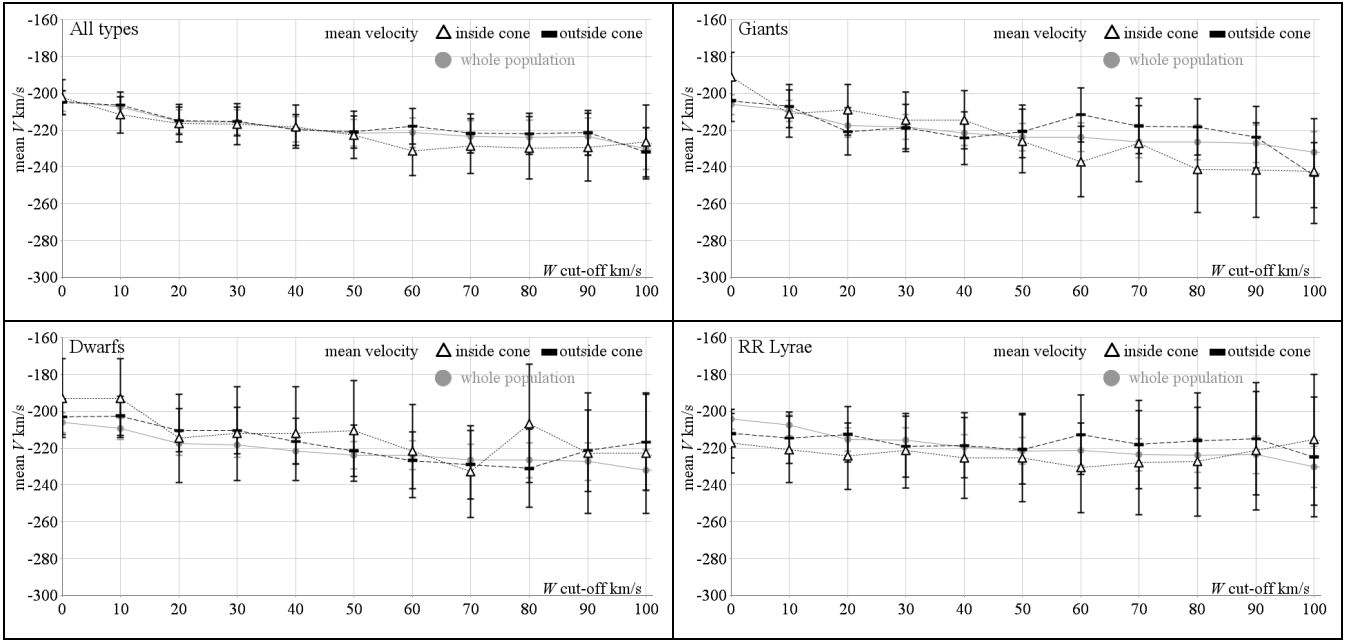

**Figure 7:** As figures 5 & 6, but with a systematic 23% cut in radial velocity, increases of distance of 7% for dwarfs and 10% for giants, and decreases in distance to RR Lyrae of 5%.

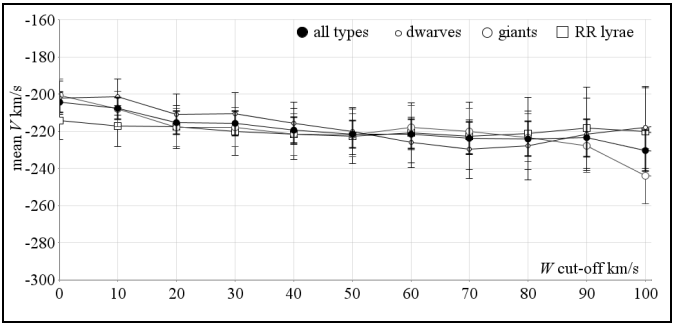

**Figure 8:** As figure 4, but with a systematic 23% cut in radial velocity, increases in distance of 7% for dwarfs and 10% for giants, and decreases in distance to RR Lyrae of 5%.

A high pass rate is found on the $W$-axis because of the Solar posion close to the galactic plane, and similarly if the result on the $V$-axis appears to indicate a that the Sun has some privileged position in the direction of orbital motion. But there are no privileged positions in the mid part of an equiangular spiral structure (an analysis of the spiral structure of the Milky Way and its relation to the observed velocity distribution is given by FA09b). The principle conclusion one can draw is that the result is not real; it shows a systematic over-statement in components of heliocentric radial velocities in the direction of orbital motion.

On account of the small population of halo stars, the calculation of Solar motion does not demonstrate an illusory component of heliocentric radial velocity at the 3σ level, or lead to a precise calculation of the orbital velocity of the Sun, but it does offer independent supporting evidence for the results of the regression test. It is to be expected that future measurements, in particular from Gaia, will provide a population of sufficient size to give a conclusive result.

After considering random and systematic measurement errors as well as possible selection bias and the observed structure of the velocity distribution, which reflects the spiral structure of the galaxy, we have not been able to find any explanation in classical physics which can produce these results. It is a prediction of general relativity that cosmological expansion does not affect spectral shifts within gravitationally bound systems, but the anomalous Pioneer blueshift (Anderson et. al., 2002) is still unexplained, and has a value remarkably close to Hubble's constant. If the Pioneer blueshift has a cosmological origin, the prediction of general relativity is violated and one must ask whether cosmological expansion might also affect other Doppler shift measurements within the Galaxy.

## 7   Conclusion

We have applied a statistical test on a population of 20 440 Hipparcos stars inside 300 pc with accurate parallaxes in the New Hipparcos Reduction, and for which radial velocities are known. We divided the population into twenty bins by colour. For each component of velocity, $U$, toward the Galactic centre, $V$, in the direction of rotation, and $W$, toward the galactic North pole, and divided each bin into four quadrants, stars approaching, stars receding, and stars whose position is in either direction along the axis. Under the null hypothesis, that *there is no systematic error in spectrographic determinations of heliocentric radial velocity*, there should be a 50-50 split of quadrants with absolute component of velocity increasing or decreasing with abs(cosθ). The split on the $V$-axes, confirms the alternate hypothesis, *heliocentric radial velocities are overstated*, with 99.95% confidence.

We have considered as possible causes of the test result: random and systematic measurement error, selection bias, truncation bias,

colour bias, orbital deviation of motion, stellar streams arising from galactic spiral structure or other cause, velocity gradients and divergences, effects of the halo and of the thick disc, effects due to extinction and other irregularities. Analysis of some of these features indicated that, if anything, a correlation opposite to that observed should be expected. We have found no explanation in classical physics which can produce these results, and indeed we believe that any reasonable classical argument can produce little deviation from the expected 50% pass rate.

In a separate test on metal-poor stars, we have found tension between calculations of the orbital velocity of the Sun and three populations of halo stars inside and outside of a cone of 60° semi-angle from the direction of rotation. Tension cannot be removed with only systematic distance adjustments.

We have concluded that there is evidence for an unmodelled element in spectrographic determinations of heliocentric radial velocity with a possible cosmological origin, affecting velocity components in the direction of orbital motion, and that this unmodelled component, rather than CDM or MOND is likely to be responsible for the apparent flatness of galaxy rotation curves. While this suggestion represents a radical departure from the usual lines of research, it should not be dismissed for that reason. As a matter of fundamental principle, science requires that all possible hypotheses are examined and eliminated through a thorough analysis of data. In the absence of direct experimental comparison between spectral shift and the motion of stars determined by direct means, in the absence also of an accepted model of quantum gravity according to which cosmological spectral shift of photon wavelengths can be shown equal to the classical calculation based on the expected behaviour of a classical electromagnetic field in general relativity, and given the failure of both CDM and MOND to answer all challenges, it is important to consider new possibilities for the interpretation of observational evidence. Simple, clear, results from individual stars must await direct measurement of radial velocity by Gaia. In the meantime we would urge astrophysicists and cosmologists to exercise caution before accepting spectrographic evidence for CDM or MOND.

**Data and Calculation**

The compiled data used in the radial velocity test can be downloaded from http://data.rqgravity.net/lsr/

The tests were carried out in Quantrix Modeler and can be downloaded from http://models.rqgravity.net/radialvelocitytests/

They can be viewed with Quantrix Viewer, which can be freely downloaded from
https://www.quantrix.com/s/viewer/InstallQuantrixViewer-3.1.exe.

A free limited period trial of Quantrix Modeler can be downloaded from http://quantrix.com/